# Pitfalls of defacing whole-head MRI: re-identification risk with diffusion models and compromised research potential

Chenyu Gao,[1*] Kaiwen Xu,[2*] Michael E. Kim,[2] Lianrui Zuo,[1] Zhiyuan Li,[1] Derek B. Archer,[3,4] Timothy J. Hohman,[3,4] Ann Zenobia Moore,[5] Luigi Ferrucci,[5] Lori L. Beason-Held,[6] Susan M. Resnick,[6] Christos Davatzikos,[7] Jerry L. Prince,[8] Bennett A. Landman[1,2,4]

[*]Authors contributed equally to this work.

[1] Department of Electrical and Computer Engineering, Vanderbilt University, Nashville, Tennessee, USA

[2] Department of Computer Science, Vanderbilt University, Nashville, Tennessee, USA

[3] Vanderbilt Memory and Alzheimer's Center, Vanderbilt University Medical Center, Nashville, Tennessee, USA

[4] Department of Neurology, Vanderbilt University Medical Center, Nashville, Tennessee, USA

[5] Translational Gerontology Branch, National Institute on Aging, National Institutes of Health, Baltimore, Maryland, USA

[6] Laboratory of Behavioral Neuroscience, National Institute on Aging, National Institutes of Health, Baltimore, Maryland, USA

[7] AI2D Center for AI and Data Science, University of Pennsylvania, Philadelphia, USA

[8] Department of Electrical and Computer Engineering, Johns Hopkins University, Baltimore, Maryland, USA

Correspondence to: Chenyu Gao

E-mail: chenyu.gao@vanderbilt.edu

Phone: 667-910-5300

Full address: 2301 Vanderbilt Pl., PO Box 351679 Station B, Nashville, TN 37235-1679

# Abstract

Defacing is often applied to head magnetic resonance image (MRI) datasets prior to public release to address privacy concerns. The alteration of facial and nearby voxels has provoked discussions about the true capability of these techniques to ensure privacy as well as their impact on downstream tasks. With advancements in deep generative models, the extent to which defacing can protect privacy is uncertain. Additionally, while the altered voxels are known to contain valuable anatomical information, their potential to support research beyond the anatomical regions directly affected by defacing remains uncertain. To evaluate these considerations, we develop a refacing pipeline that recovers faces in defaced head MRIs using cascaded diffusion probabilistic models (DPMs). The DPMs are trained on images from 180 subjects and tested on images from 484 unseen subjects, 469 of whom are from a different dataset. To assess whether the altered voxels in defacing contain universally useful information, we also predict computed tomography (CT)-derived skeletal muscle radiodensity from facial voxels in both defaced and original MRIs. The results show that DPMs can generate high-fidelity faces that resemble the original faces from defaced images, with surface distances to the original faces significantly smaller than those of a population average face ($p < 0.05$). This performance also generalizes well to previously unseen datasets. For skeletal muscle radiodensity predictions, using defaced images results in significantly weaker Spearman's rank correlation coefficients compared to using original images ($p \leq 10^{-4}$). For shin muscle, the correlation is statistically significant ($p < 0.05$) when using original images but not statistically significant ($p > 0.05$) when any defacing method is applied, suggesting that defacing might not only fail to protect privacy but also eliminate valuable information. We advocate two solutions for data sharing that comply with privacy: 1) share skull-stripped images along with measurements of facial and cranial features extracted before skull-stripping for public access, while acknowledging that this approach inherently compromises many research potentials; or 2) share the unaltered images with privacy enforced through policy restrictions.

# 1. Introduction

The practice of defacing whole-head MRI has become increasingly widespread. This trend is driven by multiple factors, including improved image quality, enhanced 3D reconstruction capabilities, more powerful facial recognition techniques, growing efforts in data sharing, and increased emphasis on data privacy due to public awareness and institutional or governmental requirements. High-resolution structural MRIs can reveal facial features that can potentially be used to identify individuals.[1,2] This type of information is protected by regulations such as the Health Insurance Portability and Accountability Act (HIPAA) in the United States and General Data Protection Regulation (GDPR) in Europe. Due to these regulations, and despite variations in specific requirements by each Institutional Review Board (IRB), more entities are applying defacing techniques to MRI datasets before sharing. For instance, the UK Biobank[3] uses *FSL_deface*[4] for T1-weighted (T1w) and T2-weighted (T2w) MRIs. Similarly, the Human Connectome Project (HCP)[5] datasets use *Face_Masking*[6] for T1w and T2w MRIs.

There are concerns about defacing MRI. Thorough discussion in the literature has brought three key issues to light. First, defacing itself can fail in two ways: it can be too conservative, failing to remove enough facial features and leaving the MRIs recognizable by facial recognition software, or it can be too aggressive, altering voxel intensities beyond the facial regions or, worse, within the brain.[7–9] Second, defacing can also cause failures in other processing pipelines, including, but not limited to image registration,[10] whole-brain segmentation,[7,11] white matter lesion segmentation,[12] and glioblastoma segmentation.[12] Third, defacing has impact on the reproducibility of analyses.[7,8,11–13] In addition to these known issues, new concerns have emerged with recent progress of deep generative models, e.g., the diffusion probabilistic models (DPMs), as these emerging technologies might be able to recover faces from defaced images. If true, this capability would challenge the rationale of applying defacing in the context of modern AI advancements. Additionally, non-brain regions in whole-head MRIs contain valuable information that could be used for various studies.[14–22] Defacing removes or alters these voxels, thereby eliminating this research potential and limiting the scope of possible scientific inquiries.

In this paper, we delve into the two emerging concerns (Figure 1). First, we show that DPMs can generate 3D high-resolution MRIs with realistic faces that resemble the original faces from defaced images, demonstrating a potential malicious privacy attack. Second, we investigate a specific example of the valuable information contained within facial voxels that are typically removed by defacing. Specifically, we perform skeletal muscle radiodensity predictions using facial voxels in original and defaced MRIs and compare the discrepancy in performance.

## 1.1 High-Resolution 3D Refacing of Defaced MRI

Refacing can be considered an inpainting task, where a generative model, trained to approximate the probability distribution of the data, imputes missing regions by sampling from the probability distribution based on the observed regions. In medical image analysis, previous studies have demonstrated the possibility to extend imaging boundaries to restore anatomical structures truncated by restricted field-of-view,[23,24] adding value to downstream clinically relevant applications as a normalization step.[25] Comparatively, the application of generative models for refacing MRI has been scarce. Abramian et al. used CycleGAN to generate refaced sagittal slices from defaced sagittal slices.[26] Similarly, Xiao et al. used a pixel-constrained convolutional neural network (CNN) for 2D sagittal refacing at a lower resolution.[27] Both of these works performed refacing in 2D, which is insufficient to pose a significant privacy threat. As the authors of these papers acknowledged, 3D generative models and reconstruction are needed for more effective refacing.[26,27] Molchanova et al., on the other hand, used a 3D conditional generative adversarial network for refacing.[28] However, their goal was to provide another defacing (de-identification) tool that replaces the original face with a generic face (similar to what *mri_reface*[8] does), rather than to demonstrate a potential malicious privacy attack. Consequently, the faces in their refaced images do not resemble the original faces closely enough to be considered a threat.

Our goal is to explore the possibility of recovering faces from defaced images using DPMs, which could potentially be used in malicious privacy attacks, rendering the practice and efforts of defacing ineffective. We input defaced MRI into cascaded DPMs to generate high-resolution 3D images with faces that resemble the original faces prior to defacing. We then measure the similarity between the refaced faces and the original faces to assess the re-identification risk.

## 1.2 Predicting Skeletal Muscle Radiodensity from Original vs. Defaced Facial Voxels

Non-brain regions in whole-head MRI have been shown to contain valuable information in various contexts. For instance, Hitomi et al. demonstrated that gadolinium leakage into ocular structures on postcontrast FLAIR MRI is associated with acute stroke.[14] Wiseman et al. showed that the axial length of the eye can be measured from MRI, providing a proxy for eye size in the absence of biometry.[15] Mi et al. quantified temporalis muscle area from MRI, providing a sarcopenia-related metric associated with oncological outcomes.[16] These examples represent only a small fraction of the anatomical and clinical insights that could be derived from non-brain regions. Other regions, such as the meninges[17], tongue[18], sinuses[19], ear canals[20], marrow[21], and subcutaneous fat[21,22], also hold significant research potential that could be compromised by defacing.

We explore the value of non-brain regions from a new perspective by predicting abdomen, thigh, and shin muscle radiodensities, measured from body CT data, from facial voxels in head MRI. Specifically, we include subjects with paired body CT and T1w MRI and apply four defacing methods to the T1w images. Using the remaining facial voxels from each type of defaced images, we train separate 3D residual neural networks (ResNet) to predict the muscle radiodensity of the abdomen, thigh, and shin, respectively. To examine the impact of information loss caused by defacing, we compare the correlation between predicted and ground truth values derived from defaced images with the correlation derived from original (non-defaced) images.

# 2. Cascaded Diffusion Models for Refacing Defaced MRI

We consider a two-stage refacing pipeline that takes in defaced MRIs and generates high-resolution 3D MRIs with faces intended to resemble the original faces. We measure the similarity between the generated images and the original images from both a face re-identification risk perspective and an image quality perspective.

## 2.1 Model Architecture

Inspired by the cascaded diffusion models described by Ho et al.[29], we use a two-stage approach to achieve high-resolution 3D image generation (Figure 2). This approach offers advantages over directly training a single 3D model in high resolution, including computational efficiency.

Our pipeline consists of two diffusion models, one for each stage, both of which are Denoising Diffusion Implicit Models (DDIM)[30] (Figure 2). The stage-1 model is 3D and imputes the missing facial voxels, conditioned on the downsampled defaced image, to generate a low-resolution refaced image. The stage-2 model is 2.5D and performs super-resolution enhancement, conditioned on the up-sampled, low-resolution refaced image from stage-1 and the defaced image. This is done in a slab-wise manner, where each slab consists of a stack of 8 axial slices. To mitigate border effects commonly observed in 2.5D approaches, adjacent slabs have 4 overlapping slices, allowing them to share anatomical context. The high-resolution slabs from stage-2 are then merged to form the complete high-resolution 3D refaced image. Our implementation of the diffusion models is adapted from Song et al.'s implementation[30], available at https://github.com/ermongroup/ddim. We slightly modify the DDIM denoising scheme to let the model to predict the ground-truth image $x_0$ at each denoising step $t$ instead of predicting the noise $\epsilon_\theta^{(t)}$. $\epsilon_\theta^{(t)}$ is estimated using the predicted $x_0$ following the first section of formula (12) in Song et al. (2021)[30]. Then, we follow formula (12) to infer the less noisy image $x_{t-1}$ for the next denoising step. This modification avoids the challenges in converging during the denoising procedure, which is observed when directly applying DDIM denoising scheme.

## 2.2 Data

Since most MRI defacing tools work for T1w MRI, we use T1w images for our experiments. We include 21, 179, and 469 subjects from the Kirby-21 [31], OASIS3 [32], and BLSA datasets [33], respectively, with one T1w image per subject. Of the 200 subjects from Kirby-21 and OASIS3, 180 were used to train the diffusion models, while 20 were reserved for internal testing. Five of these internal testing subjects were excluded because their faces were already not visible even before defacing. The 469 BLSA subjects, whose faces were visible, serve as an external testing set. We apply *FSL_deface*[4], *MRI_Deface*[34], *Pydeface*[35], and *Quickshear*[36] to the T1w images. In total, there are 720 (180 subjects * 4 defacing methods) pairs of training data.

To examine whether the generated faces are any better than a population average face, regarding the similarity with the original faces, we run *mri_reface*[8], which replaces the original face with a population average face linearly aligned to the original face.

IRB of Vanderbilt University waived ethical approval for de-identified access of the human subject data.

## 2.3 Similarity Metrics

To quantitatively measure the similarity between faces in two images, we use the mean absolute surface distance (MASD). First, we generate a binary mask of the head in each image using Otsu's thresholding[37] followed by morphological operations. To focus the comparison on the face, we crop out 10 axial slices at the inferior to exclude the neck and the posterior half of the head to exclude the back of the head. Using these binary masks, we apply the marching cubes algorithm to extract a mesh of vertices representing the surface of each face. We then use a K-dimensional tree (KDTree)[38,39] to query the nearest neighbor on the surface of one face for every vertex on the other face and compute the absolute distance. The MASD between the two faces is obtained by averaging these distances.

To measure the similarity of two images from an image quality perspective, we report the structural similarity index measure (SSIM) and peak signal-to-noise ratio (PSNR). SSIM and PSNR are computed for two regions: 1) the whole head, defined by the binary mask of the head in the original image; and 2) the area that has been changed (Supplementary Material), specifically the intersection of areas removed by defacing, altered by *mri_reface*, and imputed by our refacing. The masks used for calculating these metrics were visually inspected for quality assurance.

# 3. Estimation of Skeletal Muscle Radiodensity from Facial Voxels

We estimate CT-derived abdomen, thigh, and shin skeletal muscle radiodensities from facial voxels in T1w MRI. We perform these estimations using either original or defaced T1w images. By comparing the results, we evaluate the impact of defacing on the accuracy of estimating body measurements from head MRI.

## 3.1 Paired MRI-CT Data

We include subjects from BLSA[33] who have paired whole-head T1w MRI and single-slice CT data of the abdomen, thigh, and shin. For the T1w images, we apply defacing methods including *FSL_deface*[4], *MRI_Deface*[34], *Pydeface*[35], and *Quickshear*[36] to generate four types of defaced images from the original images. As a baseline reference, representing the most aggressive defacing where all facial voxels are removed, we apply skull-stripping using segmentation masks produced by SLANT-TICV[40] to obtain skull-stripped images from the original images. Each original T1w image is affinely registered to the NMRI225 template[41], a T1w MRI template with a large field of view that covers the whole head. The same transformation is applied to the corresponding defaced and skull-stripped images to align them with the registered original image. To focus solely on facial voxels, we crop out the posterior two-thirds of the head to exclude the back of the head and mask out the brain to further isolate the facial voxels. Examples are available in the Supplementary Materials. The intensities of the remaining voxels in the T1w image are normalized by subtracting the mean and dividing by the standard deviation. For the single-slice CT images of the abdomen, thigh, and shin, we use deep learning-based methods to segment the muscle in each body part.[42–44] We then calculate the mean voxel intensity within the segmented muscle mask for each body part to quantify the muscle radiodensity (in Hounsfield Units) of the abdomen, thigh, and shin, respectively.

We perform quality assurance by visually inspecting the processed data and exclude MRI-CT pairs where any processing step failed, ensuring that all groups (image types) have the same number of valid data points for the regression experiments. In the end, we have 948 BLSA subjects with paired MRI and abdomen CT data. These subjects are split into 616 for training, 142 for validation, and 190 for testing. Additionally, we have 990 BLSA subjects with paired MRI and thigh and shin CT data, split into 643 for training, 148 for validation, and 199 for testing.

IRB of Vanderbilt University waived ethical approval for de-identified access of the human subject data.

## 3.2 Regressing Out Age and Sex

Age and sex are two confounding factors that are strongly correlated with skeletal muscle radiodensity. Since our goal is to investigate whether facial voxels in T1w images contain information specific to abdomen, thigh, and shin muscle radiodensity, we regress these two confounding factors out of the muscle radiodensity measurements. Specifically, we fit three linear mixed-effects models[45], one for each body part (abdomen, thigh, or shin), using the formula:

$$y_{i,j} = \beta_0 + \beta_1 \times \text{Age}_{i,j} + \beta_2 \times \text{Sex}_i + r_i + \varepsilon_{i,j} \qquad (1)$$

where $y_{i,j}$ is the muscle radiodensity of subject $i$ at visit $j$, $\text{Age}_{i,j}$ is the age of subject $i$ at visit $j$, $\text{Sex}_i$ is the sex of subject $i$, $r_i$ represents the random intercept for each subject, and $\varepsilon_{i,j}$ is the residual error of the linear mixed-effects model. We then subtract the predicted muscle radiodensity $\hat{y}_{i,j}$ by age and sex from the observed muscle radiodensity to obtain the residuals, using the equation:

$$y_{i,j} - \hat{y}_{i,j} = y_{i,j} - \beta_0 - \beta_1 \times \text{Age}_{i,j} - \beta_2 \times \text{Sex}_i \qquad (2)$$

These residuals will be the target of our regression models, which are introduced in the next section.

## 3.3 Regression Model

We use 3D residual neural networks (ResNet)[46], specifically ResNet10, initialized with weights pretrained on large medical image datasets[47]. The input to the ResNet is one of six types of cropped 3D images containing only facial voxels. Each image is intensity-normalized before being fed into the ResNet. The target of the ResNet is the muscle radiodensity in one of the body parts (abdomen, thigh, or shin) with age and sex regressed out using equation (2). The values are normalized using the mean and standard deviation computed from the training set. For each input-target pair, we train a separate ResNet. In total, there are 18 (6 types of images * 3 body parts) ResNets. All ResNets are trained using the same hyperparameters. To mitigate the models' bias towards subjects with more samples, the probability of selecting each sample is normalized based on the number of samples available for each subject.

## 4. Results

Figure 3 shows a testing subject from Kirby-21 whose face was removed by *MRI_Deface* and subsequently recovered by our DPMs. The DPMs-refaced face closely resembles the original face, based on visual inspection, whereas the linearly aligned population average face appears less similar to the original face. This visual observation is consistent with the MASD measurements: for DPMs-refaced face, the MASD to the original face is 0.363 mm, compared to 0.792mm for the linearly aligned population average face. White patches were placed over the eyes in the screenshots of the renderings to obscure subject identity, but all computations were performed without such obscuring. For the same subject, we present 3D renderings of the images refaced from each type of defaced image by our DPMs, ordered by ascending MASD to the original face (Figure 4). The MASD generally aligns with the perceptual dissimilarity between the two faces. On skull-stripped images, the DPMs (retrained on skull-stripped images) fail to reconstruct faces that resemble the original ones (Figure 5 and 6).

On the internal testing set of 15 subjects, MASDs between DPMs-refaced faces and original faces are significantly smaller compared to MASDs between population average faces and original faces (Table 1). Among the DPMs-refaced faces, those generated from *MRI_Deface*-defaced images have the smallest MASD (0.34 mm), while those generated from *Quickshear*-defaced images have the largest MASD (0.63 mm). Similarly, on the external testing set of 469 subjects, MASDs between DPMs-refaced faces and original faces are significantly smaller compared to MASDs between population average faces and original faces. Among the DPMs-refaced faces, those generated from *MRI_Deface*-defaced images have the smallest MASD (0.38 mm), while those generated from *Pydeface*-defaced images have the largest MASD (0.67 mm). In general, DPMs-refaced images have higher PSNR and SSIM in the whole head area, but lower PSNR and SSIM in the facial area removed by defacing, compared to *mri_reface* processed images (Tables 2 and 3).

For all three body parts, defacing results in significantly lower Spearman's rank correlation coefficients between the predicted and ground truth values of muscle radiodensity ($p \leq 10^{-4}$) (Figure 7). When predicting abdomen muscle radiodensity, the correlation is significant (p < 0.05, as shown by the 95% confidence intervals) when using original images, but not significant (p > 0.05) when using defaced images by *FSL_deface*, *MRI_Deface*, and *Pydeface*. When predicting shin muscle radiodensity, the correlation is significant only when using original images (p < 0.05), while the use of any defacing method results in nonsignificant correlations (p > 0.05).

# 5. Discussion

## 5.1 The cascaded DPMs can generate MRIs with high-fidelity faces from defaced MRIs

Our refacing pipeline based on DPMs can generate high-resolution 3D T1w MRIs with faces that resemble the original faces, from defaced images. We used MASD as a metric to quantify the similarity between two faces. This metric generally aligns with our perceptual impression of whether two faces belong to the same person (Figure 4). The DPMs-refaced faces have a smaller MASD to the original faces compared to the linearly aligned population average face. The mechanism behind this capability is twofold. First, the DPMs learn a statistical prior of craniofacial anatomy from the training set. This process is related to the known ability of generative models to "memorize" and replicate features from their training data, a significant privacy concern in itself.[48] Second, and more importantly, the model uses the remaining, non-defaced portions of the skull and surrounding tissue as a strong condition to guide the generative process. This transforms the task into a conditional inpainting problem, where the learned anatomical prior is constrained by the unique anatomy of the individual subject. Therefore, the DPMs leverage the underlying anatomical information present in the MRIs, allowing the model to reconstruct the defaced regions in a way that is more personalized and accurate, reflecting individual anatomical features rather than just an averaged representation. Additionally, we found that the more aggressive the defacing, the more challenging the refacing becomes, as indicated by a larger MASD to the original faces (Figure 4 and Table 1). On the skull-stripped images, the DPMs fail to reconstruct the faces (Figure 5 and 6), providing strong evidence for this conditional mechanism, as the necessary anatomical context is absent. However, this does not necessarily make skull-stripping an ideal solution, because skull-stripping removes non-brain voxels that contain valuable information, as we will discuss in the next section.

On the other hand, the DPMs-refaced faces do not achieve higher (and often even lower) SSIM and PSNR compared to the population average face within the facial voxels (Tables 2 and 3). We attribute this to two main factors. First, the population average face applied by *mri_reface* was registered to the original faces, whereas the DPMs generate the face without any access to the original faces during inference. Consequently, the linearly aligned population average face aligns the internal structure of the face better compared to the DPMs-refaced faces. Second, the DPMs were trained on a relatively small dataset of only 180 subjects. It is reasonable to expect that using larger datasets that encompass a diverse range of cohorts for training would lead to higher SSIM and PSNR values for the refaced images. However, we refrained from using large-scale, multi-site datasets in this refacing-related project to avoid potential violations of the data use agreements.

## 5.2 Facial voxels removed by defacing contain valuable information

As mentioned in Section 1.2, studies have shown the value of non-brain regions in whole-head MRIs for various applications. These studies provide examples where defacing may compromise the feasibility of certain research, as the regions of interest are corrupted by defacing. To further illustrate the impact of defacing on studies beyond the regions directly altered, we explore the connections between MRI facial features and overall body compositions. We specifically focus on muscle radiodensity derived from CT as a metric of interest, as previous studies have shown its association with muscle fat infiltration, which is predictive of various health outcomes, including metabolic diseases and overall mortality.[25] Our findings indicate that predicting abdomen, thigh, and shin muscle radiodensities from facial voxels is feasible using original (pre-defaced) MRIs, as evidenced by significant correlations between the predicted and ground truth values for all body parts. However, when using defaced images, the correlations captured by the prediction models are significantly weaker and, in many cases, not statistically significant.

## 5.3 Future studies

First, to assess whether a lower MASD corresponds to a higher success rate of facial re-identification, a future study involving face recognition is needed, for example, using a standard face recognition algorithm (e.g., ArcFace[49]). Such a study will require approval from the IRB and informed consent from the human subjects to ensure ethical compliance. Second, we used DPMs as an example to demonstrate a potential malicious privacy attack through refacing because of their performance in high-fidelity image generation.[29,30] Other generative models may also be capable for similar purposes; however, comparing their performances is beyond the scope of this paper. Third, de-identification through defacing is just one of the privacy-preserving approaches for neuroimages. There are other techniques, such as encryption,[50] summary statistics,[51] decentralized computation,[52] and federated learning,[53] that could potentially be integrated to provide a more comprehensive privacy protection solution. Importantly, we note that there is no "once and for all" solution, and strategies should adapt as (both protectors' and attackers') techniques evolve.[54]

# 6. Conclusion

These concerning results indicate that defacing might not only fail to protect privacy in the face of DPMs but also eliminate valuable information, thereby compromising research potential. We advocate two solutions for data sharing in compliance with privacy: 1) share skull-stripped images along with measurements of facial and cranial features extracted before skull-stripping (e.g., intracranial volume) for public access, while acknowledging that this approach inherently compromises many research potentials; or 2) share the unaltered images with privacy enforced through the use of policy restrictions.

# Data Sharing Statement

Our models will not be released. Code is available upon request and agreement not to use or disseminate it for re-identification purposes.

# Acknowledgments

This project is funded by NIH grant 1R01EB017230, U24AG074855, K01-AG073584. The Vanderbilt Institute for Clinical and Translational Research (VICTR) is funded by the National Center for Advancing Translational Sciences (NCATS) Clinical Translational Science Award (CTSA) Program, Award Number 5UL1TR002243-03. The content is solely the responsibility of the authors and does not necessarily represent the official views of the NIH. This work was conducted in part using the resources of the Advanced Computing Center for Research and Education at Vanderbilt University, Nashville, TN. The BLSA is supported by the Intramural Research Program, National Institute on Aging, NIH.

# Declaration of Competing Interests

The authors report no competing interests.

# Declaration of generative AI and AI-assisted technologies in the writing process

During the preparation of this work the author, Chenyu Gao, used GPT-4o in order to check grammar. After using this tool/service, the author reviewed and edited the content as needed and takes full responsibility for the content of the publication.

# Author Contributions

**Chenyu Gao:** Conceptualization, Methodology, Software, Validation, Formal analysis, Investigation, Data Curation, Writing - Original Draft, Writing - Review & Editing, Visualization. **Kaiwen Xu:** Conceptualization, Methodology, Software, Validation, Writing - Review & Editing. **Michael E. Kim:** Validation, Data Curation, Writing - Review & Editing. **Lianrui Zuo:** Validation, Writing - Review & Editing. **Zhiyuan Li:** Writing - Review & Editing. **Derek B. Archer:** Writing - Review & Editing, Funding acquisition. **Timothy J. Hohman:** Writing - Review & Editing,

Funding acquisition. **Ann Zenobia Moore:** Resources, Writing - Review & Editing. **Luigi Ferrucci:** Resources, Writing - Review & Editing. **Lori L. Beason-Held:** Resources, Writing - Review & Editing. **Susan M. Resnick:** Resources, Writing - Review & Editing. **Christos Davatzikos:** Writing - Review & Editing. **Jerry L. Prince:** Writing - Review & Editing. **Bennett A. Landman:** Conceptualization, Validation, Resources, Writing - Review & Editing, Supervision, Project administration, Funding acquisition.

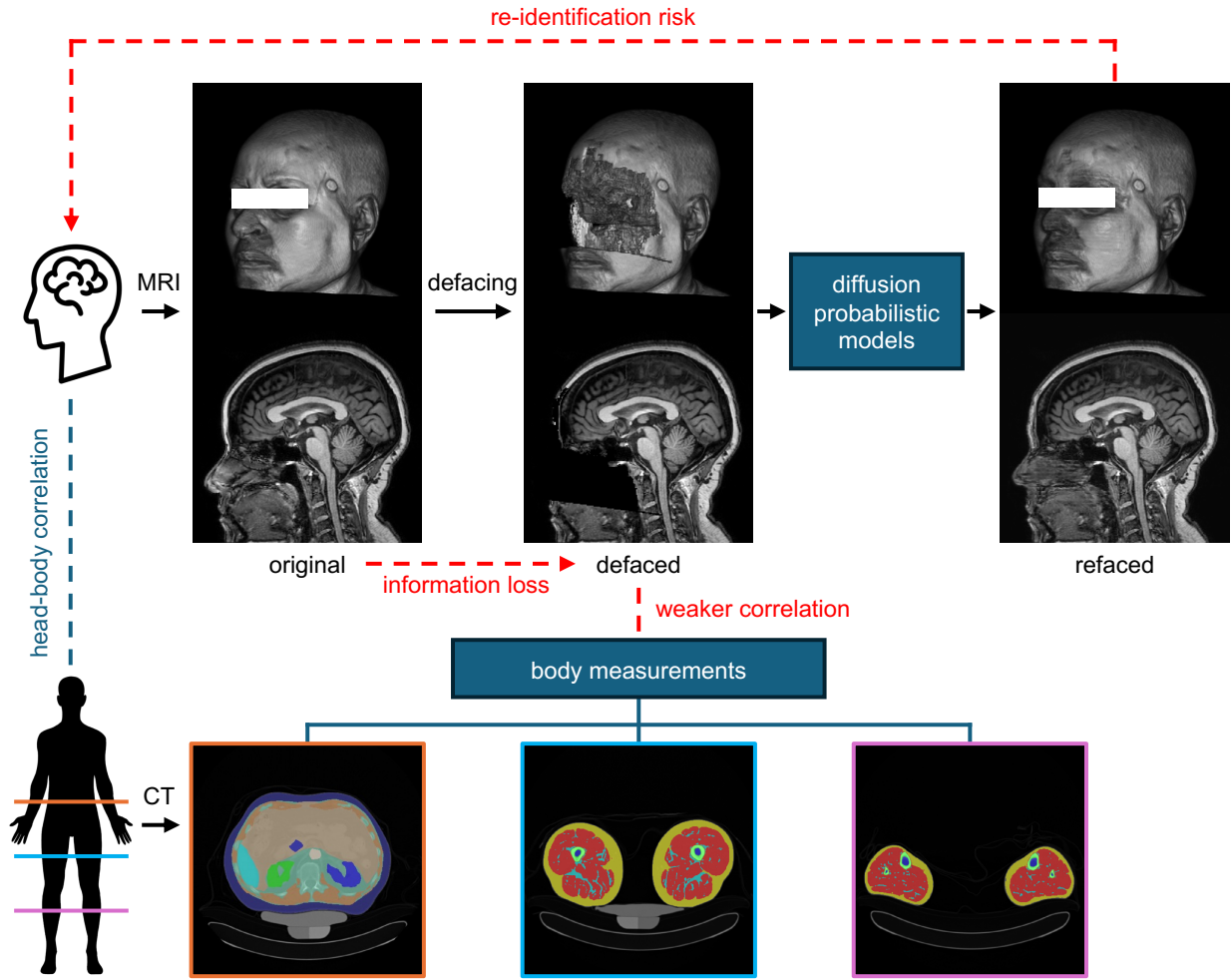

**Figure 1.** There are pitfalls of defacing, a technique used to alter facial voxels in whole-head MRIs to protect privacy. First, with deep generative models such as diffusion probabilistic models, it is possible to synthesize MRIs with realistic faces, which closely resemble the original faces, from defaced MRIs. This capability poses a re-identification risk, thus questioning the efficacy of defacing in protecting privacy. Second, facial and other non-brain voxels in whole-head MRIs contain valuable anatomical information. For instance, this information could be used to study correlations between head and body measurements using paired head MRI and body CT data. The alteration of these voxels results in information loss, thereby compromising such research potentials. The experiments in this paper are designed to showcase these two pitfalls.

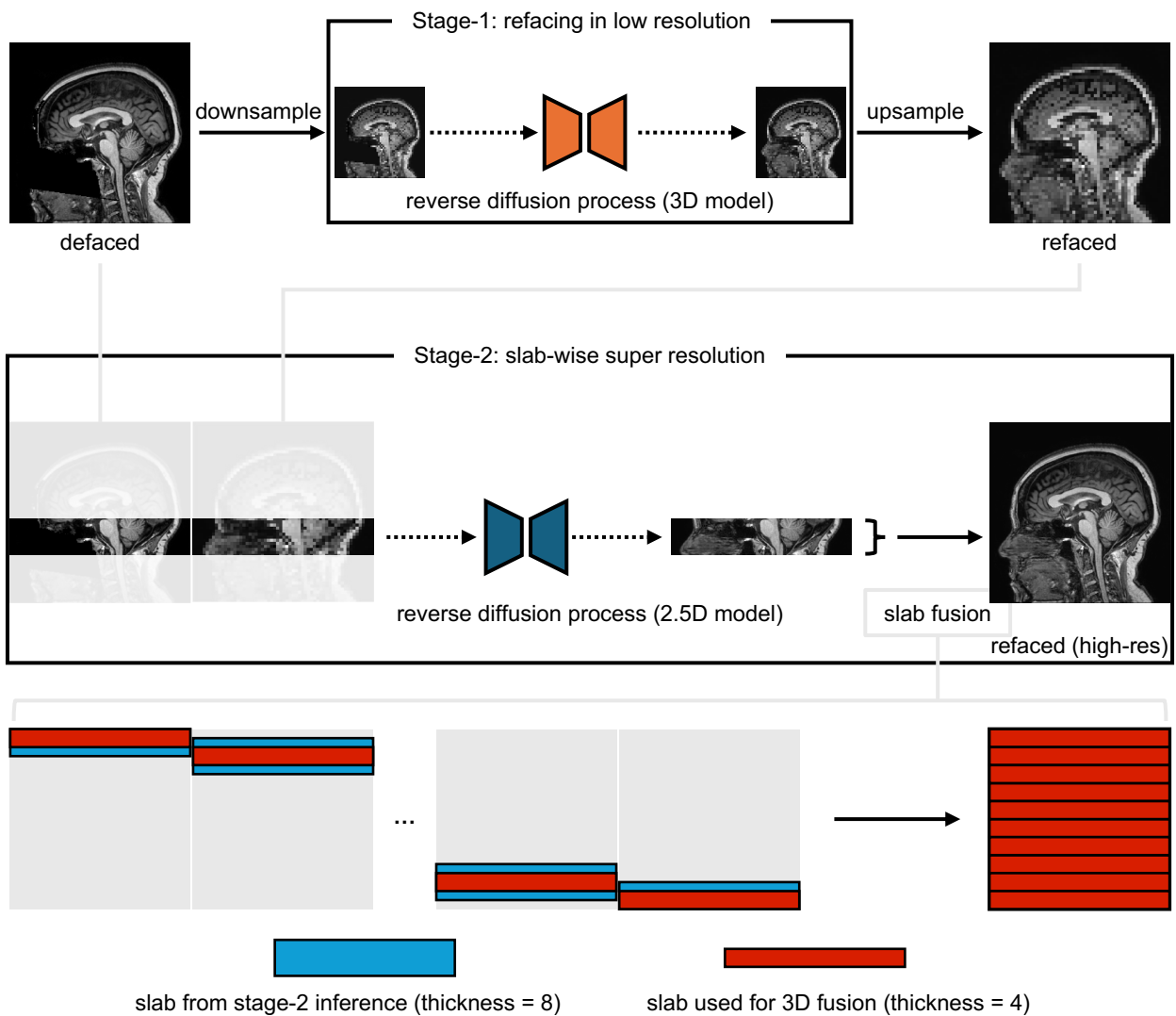

**Figure 2.** Given a defaced T1-weighted (T1w) MRI, our refacing pipeline generates a T1w image with a face. The pipeline consists of two stages. In stage 1, a 3D diffusion model, conditioned on the downsampled defaced image, generates a low-resolution refaced image. In stage 2, a 2.5D diffusion model, conditioned on the high-resolution defaced image and the up-sampled low-resolution refaced image from stage 1, generates a high-resolution refaced image. This is done in a slab-by-slab manner, where each slab consists of a stack of axial slices. To mitigate border effects, adjacent slabs have overlapping slices to share anatomical context. Finally, the high-resolution slabs produced by stage 2 are merged to form the complete high-resolution 3D refaced image.

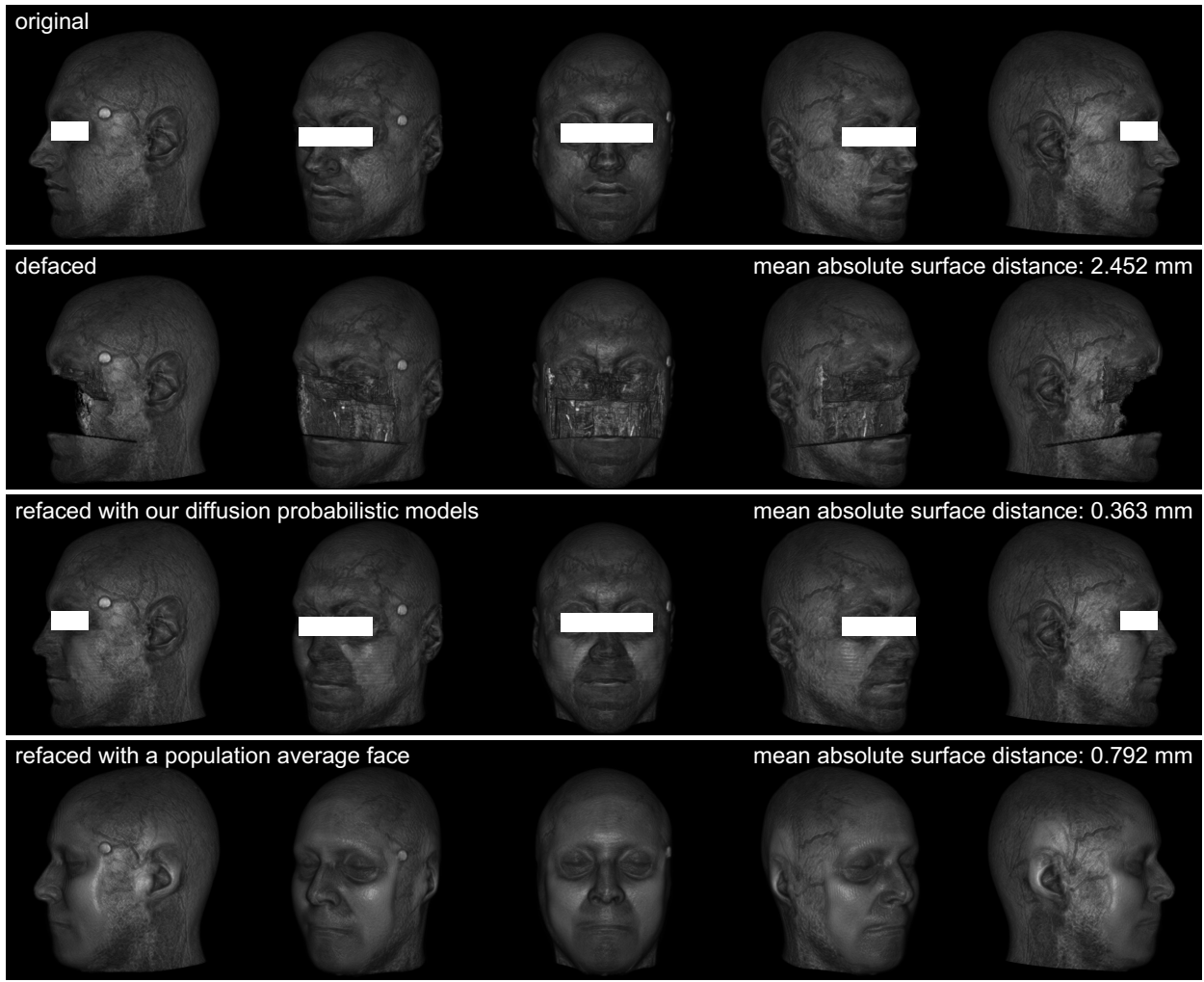

**Figure 3.** For an example subject, 3D renderings of the faces in the original image, image defaced by *MRI_Deface*, image refaced from the defaced image by our DPMs, and image processed with *mri_reface* (which replaces the original face with a population average face) are presented in each row. The MASD is computed between the original face and the face in each image type. A smaller distance corresponds to a higher similarity to the original face.

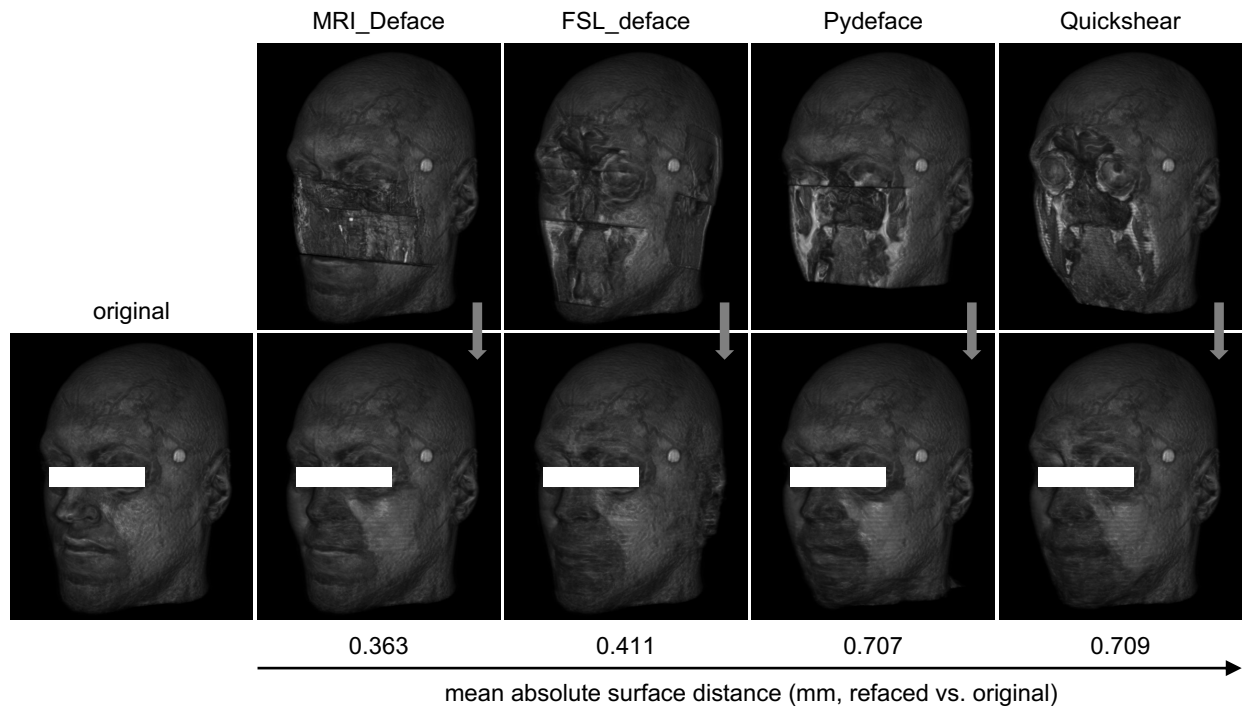

**Figure 4.** For the same subject in Figure 3, 3D renderings of the faces in the DPMs-refaced images generated from each type of defaced image are presented alongside the original face. The images are ordered by the MASD to the original face. As the distance increases, the face appears perceptually more different from the original face.

**Table 1.** Mean absolute surface distance between the original face and the altered (or synthetic) face.

| Testing set | Defaced by | Mean absolute surface distance to the original face (mm)[†] | | | p-value[‡] |
| --- | --- | --- | --- | --- | --- |
| | | Defaced | Refaced with a population average[*‡] | Refaced with DPMs[‡] | |
| Internal (N=15) | FSL_deface | 3.65 [3.27, 4.05] | 0.81 [0.69, 0.93] | 0.60 [0.53, 0.67] | 0.026 |
| | MRI_Deface | 2.35 [1.91, 2.83] | 0.81 [0.69, 0.93] | 0.34 [0.28, 0.40] | 0.001 |
| | Pydeface | 6.01 [5.69, 6.33] | 0.81 [0.69, 0.93] | 0.61 [0.53, 0.68] | 0.018 |
| | Quickshear | 5.54 [4.89, 6.20] | 0.81 [0.69, 0.93] | 0.63 [0.55, 0.71] | 0.030 |
| External (N=469) | FSL_deface | 2.97 [2.91, 3.03] | 0.71 [0.68, 0.73] | 0.63 [0.61, 0.64] | <<0.001 |
| | MRI_Deface | 2.19 [2.16, 2.22] | 0.71 [0.68, 0.73] | 0.38 [0.36, 0.39] | <<0.001 |
| | Pydeface | 6.01 [5.95, 6.06] | 0.71 [0.68, 0.73] | 0.67 [0.65, 0.69] | <<0.001 |
| | Quickshear | 5.01 [4.85, 5.18] | 0.71 [0.68, 0.73] | 0.62 [0.60, 0.64] | <<0.001 |

[*]: the population average face (linearly aligned with the original face) is applied by mri_reface. [†]: mean value and 95% confidence intervals (in the square bracket) are estimated with bootstrapping (n=1000). [‡]: Wilcoxon signed-rank test is performed between population average derived distances and DPMs derived distances.

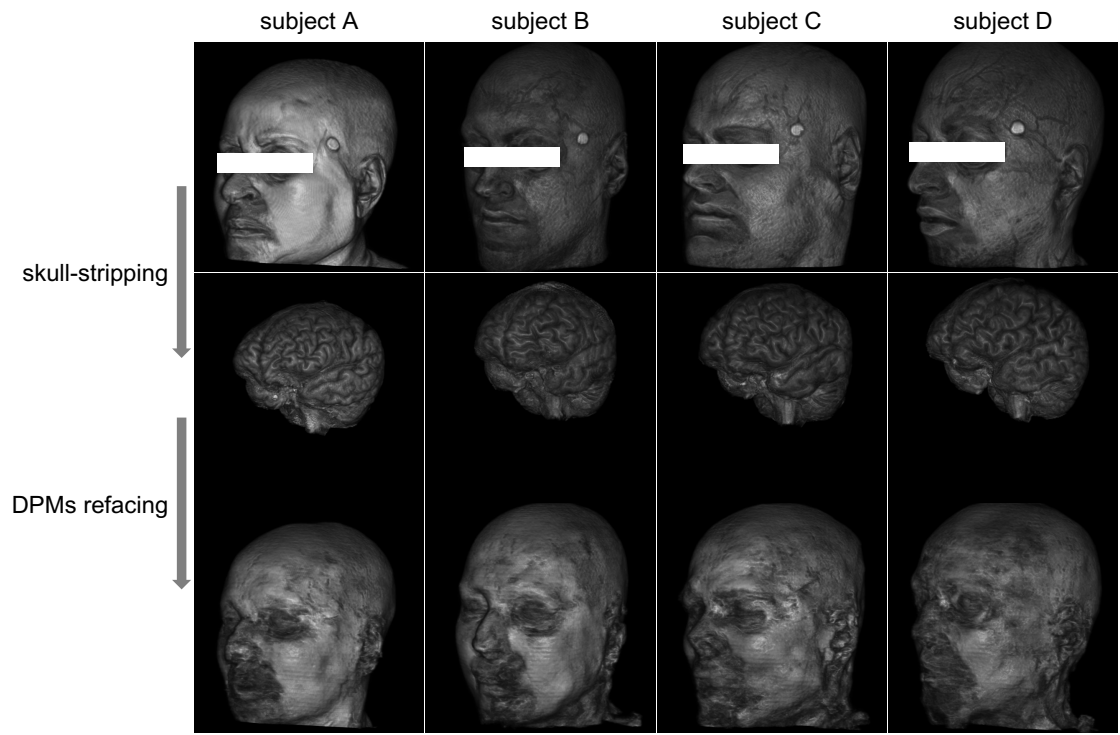

**Figure 5.** From skull-stripped images, the diffusion probabilistic models (DPMs) failed to generate faces that resemble the original faces.

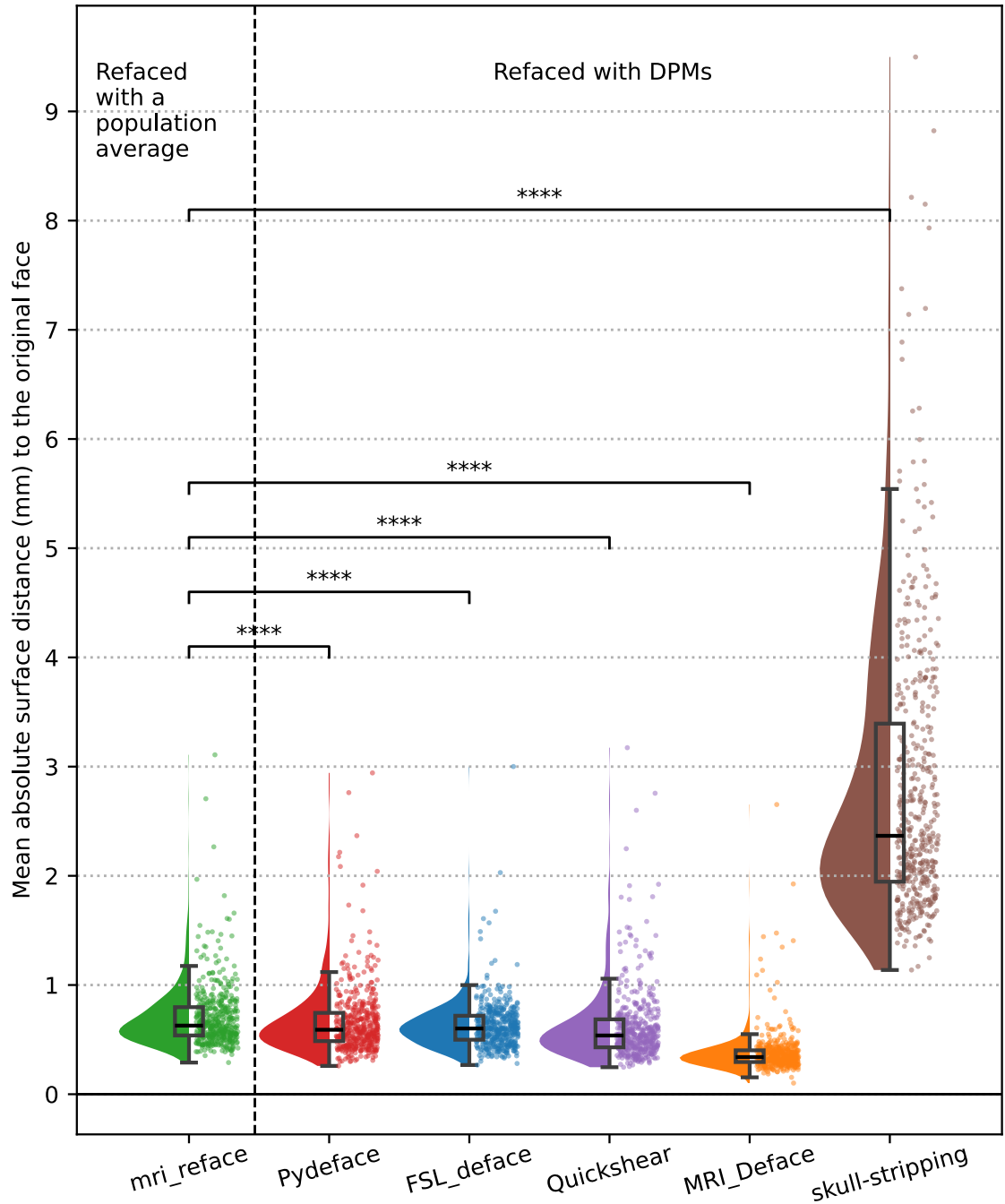

**Figure 6.** On the external testing set (N=469), faces generated by our DPMs from images defaced using Pydeface, FSL_deface, Quickshear, and MRI_Deface show significantly lower mean absolute surface distances (MASDs) to the original faces, compared to the linearly aligned population average face produced by mri_reface, indicating higher facial similarity. In contrast, faces generated by DPMs from skull-stripped images show significantly higher MASDs, indicating lower facial similarity. Statistical significance was assessed using the Wilcoxon signed-rank test; "****" indicates p ⩽ 0.0001.

**Table 2.** Peak signal-to-noise ratio (PSNR) of defaced, mri_reface, and DPMs-refaced images.

| Testing set | Defaced by | Peak signal-to-noise ratio (PSNR)[†] | | | | | |
|---|---|---|---|---|---|---|---|
| | | whole head | | | face (area removed by defacing) | | |
| | | defaced | mri_reface | DPMs-refaced | defaced | mri_reface | DPMs-refaced |
| Internal (N=15) | FSL_deface | 15.40 [14.87, 15.91] | 20.23 [19.78, 20.68] | **22.44 [21.80, 23.05]** | 4.69 [4.48, 4.90] | **12.09 [11.76, 12.43]** | 11.81 [11.57, 12.06] |
| | MRI_Deface | 13.96 [13.21, 14.63] | 20.23 [19.78, 20.68] | **21.11 [20.49, 21.65]** | 4.63 [4.28, 4.94] | **12.90 [12.48, 13.36]** | 11.29 [11.01, 11.60] |
| | Pydeface | 14.34 [13.59, 15.08] | 20.23 [19.78, 20.68] | **20.89 [20.18, 21.55]** | 5.33 [4.99, 5.68] | **12.38 [12.01, 12.79]** | 12.01 [11.69, 12.35] |
| | Quickshear | 17.05 [16.34, 17.73] | 20.23 [19.78, 20.68] | **23.07 [22.44, 23.71]** | 5.58 [5.25, 5.91] | 11.94 [11.60, 12.28] | **11.96 [11.67, 12.25]** |
| External (N=469) | FSL_deface | 16.07 [15.97, 16.17] | 20.39 [20.30, 20.47] | **23.02 [22.91, 23.13]** | 4.51 [4.47, 4.54] | **11.43 [11.37, 11.49]** | 11.34 [11.29, 11.38] |
| | MRI_Deface | 13.48 [13.44, 13.52] | 20.39 [20.30, 20.47] | **20.73 [20.68, 20.78]** | 4.21 [4.17, 4.24] | **12.62 [12.55, 12.68]** | 11.25 [11.20, 11.29] |
| | Pydeface | 13.38 [13.28, 13.49] | **20.39 [20.30, 20.47]** | 19.16 [19.02, 19.30] | 4.88 [4.84, 4.92] | **12.29 [12.22, 12.35]** | 11.54 [11.49, 11.59] |
| | Quickshear | 16.82 [16.67, 16.97] | 20.39 [20.30, 20.47] | **22.96 [22.79, 23.13]** | 4.92 [4.88, 4.97] | **11.67 [11.60, 11.74]** | 11.41 [11.35, 11.47] |

[†]: mean value and 95% confidence intervals (in the square bracket) are estimated with bootstrapping (n=1000).

**Table 3.** Structural similarity index measure (SSIM) of defaced, *mri_reface*, DPMs-refaced images.

| Testing set | Defaced by | Structural similarity index measure (SSIM)[†] | | | | | |
|---|---|---|---|---|---|---|---|
| | | whole head | | | face (area removed by defacing) | | |
| | | defaced | mri_reface | DPMs-refaced | defaced | mri_reface | DPMs-refaced |
| Internal (N=15) | FSL_deface | 0.92 [0.91, 0.93] | 0.90 [0.89, 0.90] | **0.93 [0.92, 0.94]** | 0.00 [0.00, 0.00] | **0.28 [0.27, 0.30]** | 0.17 [0.16, 0.18] |
| | MRI_Deface | 0.87 [0.86, 0.88] | 0.90 [0.89, 0.90] | **0.90 [0.89, 0.91]** | 0.00 [0.00, 0.00] | **0.38 [0.36, 0.39]** | 0.20 [0.19, 0.21] |
| | Pydeface | 0.87 [0.86, 0.88] | 0.90 [0.89, 0.90] | **0.90 [0.89, 0.91]** | 0.02 [0.02, 0.02] | **0.24 [0.21, 0.26]** | 0.19 [0.18, 0.20] |
| | Quickshear | 0.92 [0.92, 0.93] | 0.90 [0.89, 0.90] | **0.93 [0.93, 0.94]** | 0.00 [0.00, 0.00] | **0.20 [0.17, 0.22]** | 0.16 [0.15, 0.17] |
| External (N=469) | FSL_deface | 0.93 [0.93, 0.93] | 0.91 [0.90, 0.91] | **0.94 [0.94, 0.94]** | 0.00 [0.00, 0.00] | **0.30 [0.30, 0.31]** | 0.19 [0.19, 0.19] |
| | MRI_Deface | 0.87 [0.87, 0.87] | **0.91 [0.90, 0.91]** | 0.90 [0.90, 0.90] | 0.01 [0.01, 0.01] | **0.40 [0.40, 0.41]** | 0.24 [0.24, 0.24] |
| | Pydeface | 0.84 [0.84, 0.84] | **0.91 [0.90, 0.91]** | 0.86 [0.86, 0.87] | 0.02 [0.02, 0.02] | **0.30 [0.29, 0.30]** | 0.21 [0.21, 0.21] |
| | Quickshear | 0.93 [0.93, 0.93] | 0.91 [0.90, 0.91] | **0.94 [0.93, 0.94]** | 0.00 [0.00, 0.00] | **0.23 [0.23, 0.24]** | 0.18 [0.17, 0.18] |

[†]: mean value and 95% confidence intervals (in the square bracket) are estimated with bootstrapping (n=1000).

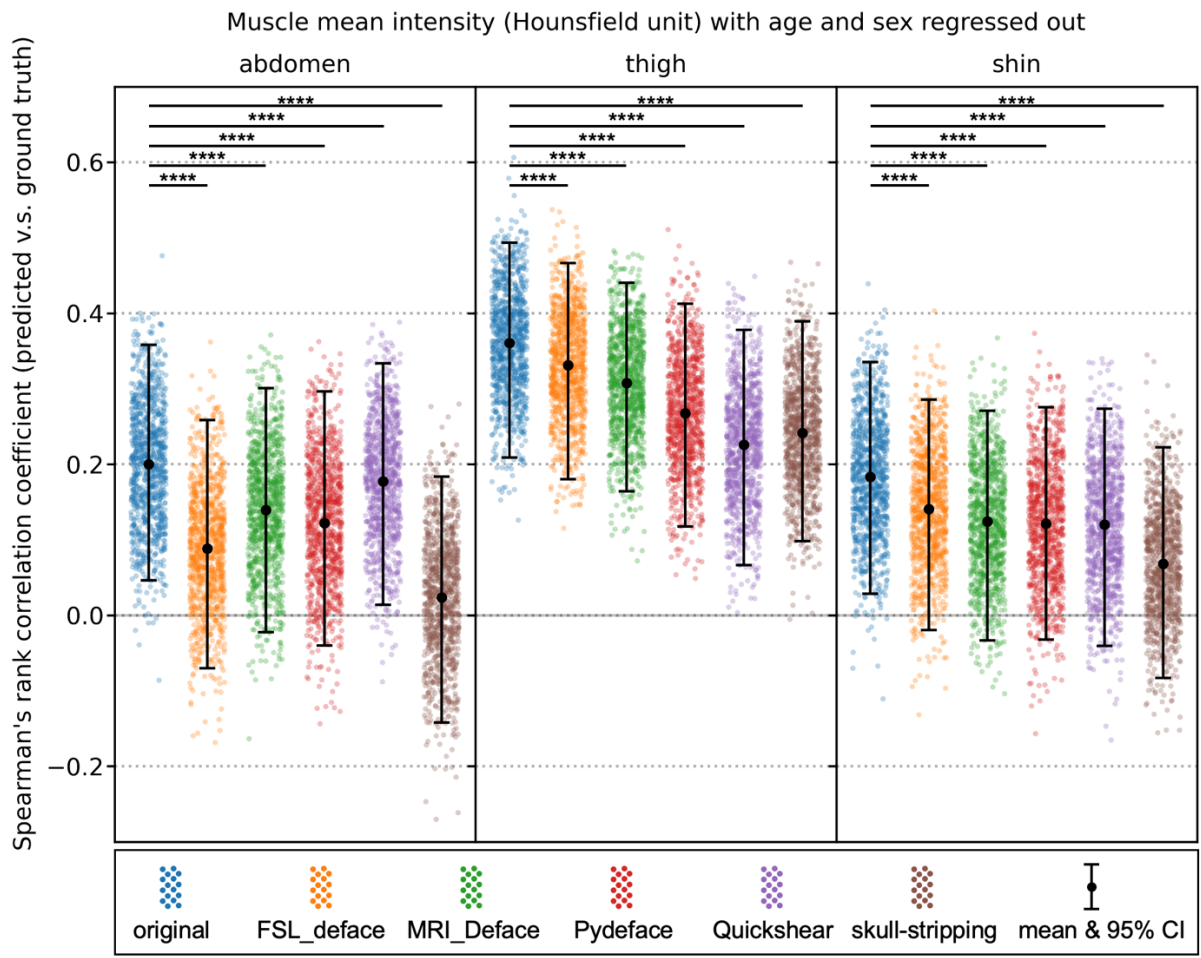

**Figure 7.** The correlation between predicted residuals of abdomen, thigh, and shin muscle radiodensity (measured from segmented ROI in CT data, in Hounsfield units, with age and sex regressed out using linear models) and ground truth values is stronger using original images compared to defaced or skull-stripped images. Examples of each image type are provided in the Supplementary Materials. The image types are arranged such that those with more facial voxels removed are placed on the right (e.g., skull-stripping), while those with fewer facial voxels removed are placed on the left (e.g., original). As more voxels are removed, the correlation between head and body becomes weaker and more difficult to capture. For instance, the correlations between predicted abdomen muscle radiodensity from FSL_deface, MRI_Deface, Pydeface, and skull-stripped images and the ground truth values are not statistically significant, as indicated by 95% confidence intervals (CI) overlapping with 0. For shin muscle, the head-body correlation is statistically significant only when using original images. Bootstrapping (n=1000) is used to estimate the mean and 95% confidence intervals of the Spearman's rank correlation coefficients. Statistical significance is indicated by "****" for p-value ≤ $10^{-4}$, based on the Wilcoxon signed-rank test.

# References

1. G SC, K KW, M TT, et al. Identification of Anonymous MRI Research Participants with Face-Recognition Software. *New England Journal of Medicine*. 2019;381(17):1684-1686. doi:10.1056/NEJMc1908881

2. Schwarz CG, Kremers WK, Arani A, et al. A face-off of MRI research sequences by their need for de-facing. *Neuroimage*. 2023;276:120199. doi:https://doi.org/10.1016/j.neuroimage.2023.120199

3. Littlejohns TJ, Holliday J, Gibson LM, et al. The UK Biobank imaging enhancement of 100,000 participants: rationale, data collection, management and future directions. *Nat Commun*. 2020;11(1):2624. doi:10.1038/s41467-020-15948-9

4. Alfaro-Almagro F, Jenkinson M, Bangerter NK, et al. Image processing and Quality Control for the first 10,000 brain imaging datasets from UK Biobank. *Neuroimage*. 2018;166:400-424. doi:https://doi.org/10.1016/j.neuroimage.2017.10.034

5. Van Essen DC, Ugurbil K, Auerbach E, et al. The Human Connectome Project: A data acquisition perspective. *Neuroimage*. 2012;62(4):2222-2231. doi:https://doi.org/10.1016/j.neuroimage.2012.02.018

6. Milchenko M, Marcus D. Obscuring Surface Anatomy in Volumetric Imaging Data. *Neuroinformatics*. 2013;11(1):65-75. doi:10.1007/s12021-012-9160-3

7. Gao C, Landman BA, Prince JL, Carass A. Reproducibility evaluation of the effects of MRI defacing on brain segmentation. *Journal of Medical Imaging*. 2023;10(6):064001. doi:10.1117/1.JMI.10.6.064001

8. Schwarz CG, Kremers WK, Wiste HJ, et al. Changing the face of neuroimaging research: Comparing a new MRI de-facing technique with popular alternatives. *Neuroimage*. 2021;231:117845. doi:https://doi.org/10.1016/j.neuroimage.2021.117845

9. Theyers AE, Zamyadi M, O'Reilly M, et al. Multisite Comparison of MRI Defacing Software Across Multiple Cohorts. *Front Psychiatry*. 2021;12. https://www.frontiersin.org/journals/psychiatry/articles/10.3389/fpsyt.2021.617997

10. Bruña R, Vaghari D, Greve A, Cooper E, Mada MO, Henson RN. Modified MRI Anonymization (De-Facing) for Improved MEG Coregistration. *Bioengineering*. 2022;9(10). doi:10.3390/bioengineering9100591

11. Gao C, Jin L, Prince JL, Carass A. Effects of defacing whole head MRI on neuroanalysis. In: *Proc.SPIE*. Vol 12032. 2022:120323W. doi:10.1117/12.2613175

12. de Sitter A, Visser M, Brouwer I, et al. Facing privacy in neuroimaging: removing facial features degrades performance of image analysis methods. *Eur Radiol*. 2020;30(2):1062-1074. doi:10.1007/s00330-019-06459-3

13. Bhalerao GV, Parekh P, Saini J, et al. Systematic evaluation of the impact of defacing on quality and volumetric assessments on T1-weighted MR-images. *Journal of Neuroradiology*. 2022;49(3):250-257. doi:https://doi.org/10.1016/j.neurad.2021.03.001

14. Hitomi E, Simpkins AN, Luby M, Latour LL, Leigh RJ, Leigh R. Blood-ocular barrier disruption in patients with acute stroke. *Neurology*. 2018;90(11):e915-e923. doi:10.1212/WNL.0000000000005123

15. Wiseman SJ, Tatham AJ, Meijboom R, et al. Measuring axial length of the eye from magnetic resonance brain imaging. *BMC Ophthalmol*. 2022;22(1):54. doi:10.1186/s12886-022-02289-y

16. Mi E, Mauricaite R, Pakzad-Shahabi L, Chen J, Ho A, Williams M. Deep learning-based quantification of temporalis muscle has prognostic value in patients with glioblastoma. *Br J Cancer*. 2022;126(2):196-203. doi:10.1038/s41416-021-01590-9

17. Duan P, Xue Y, Han S, et al. Rapid Brain Meninges Surface Reconstruction with Layer Topology Guarantee. In: *2023 IEEE 20th International Symposium on Biomedical Imaging (ISBI)*. 2023:1-5. doi:10.1109/ISBI53787.2023.10230668

18. Woo J, Murano EZ, Stone M, Prince JL. Reconstruction of High-Resolution Tongue Volumes From MRI. *IEEE Trans Biomed Eng*. 2012;59(12):3511-3524. doi:10.1109/TBME.2012.2218246

19. Chen Y, Wang X, Li L, Li W, Xian J. Differential diagnosis of sinonasal extranodal NK/T cell lymphoma and diffuse large B cell lymphoma on MRI. *Neuroradiology*. 2020;62(9):1149-1155. doi:10.1007/s00234-020-02471-3

20. Hennig L, Krüger M, Bülow R, et al. Morphology and anatomical variability of the external auditory canal: A population-based MRI study. *Annals of Anatomy - Anatomischer Anzeiger*. 2025;257:152319. doi:https://doi.org/10.1016/j.aanat.2024.152319

21. Okamoto K, Ito J, Ishikawa K, Sakai K, Tokiguchi S. Change in signal intensity on MRI of fat in the head of markedly emaciated patients. *Neuroradiology*. 2001;43(2):134-138. doi:10.1007/s002340000453

22. Hiraka T, Sugai Y, Konno Y, et al. Evaluation of the extracranial "multifocal arcuate sign," a novel MRI finding for the diagnosis of giant cell arteritis, on STIR and contrast-enhanced T1-weighted images. *BMC Med Imaging*. 2024;24(1):132. doi:10.1186/s12880-024-01314-4

23. Xu K, Li T, Khan MS, et al. Body composition assessment with limited field-of-view computed tomography: A semantic image extension perspective. *Med Image Anal*. 2023;88:102852. doi:https://doi.org/10.1016/j.media.2023.102852

24. Gao C, Bao S, Kim ME, et al. Field-of-view extension for brain diffusion MRI via deep generative models. *Journal of Medical Imaging*. 2024;11(4):044008. doi:10.1117/1.JMI.11.4.044008

25. Xu K, Khan MS, Li TZ, et al. AI Body Composition in Lung Cancer Screening: Added Value Beyond Lung Cancer Detection. *Radiology*. 2023;308(1):e222937. doi:10.1148/radiol.222937

26. Abramian D, Eklund A. Refacing: Reconstructing Anonymized Facial Features Using GANS. In: *2019 IEEE 16th International Symposium on Biomedical Imaging (ISBI 2019)*. 2019:1104-1108. doi:10.1109/ISBI.2019.8759515

27. Xiao Y, Ashbee W, Calhoun VD, Plis S. Refacing Defaced MRI with PixelCNN. In: *2022 International Joint Conference on Neural Networks (IJCNN)*. 2022:1-7. doi:10.1109/IJCNN55064.2022.9891937

28. Molchanova N, Maréchal B, Thiran JP, et al. Fast refacing of MR images with a generative neural network lowers re-identification risk and preserves volumetric consistency. *Hum Brain Mapp*. 2024;45(9):e26721. doi:https://doi.org/10.1002/hbm.26721

29. Ho J, Saharia C, Chan W, Fleet DJ, Norouzi M, Salimans T. Cascaded Diffusion Models for High Fidelity Image Generation. *Journal of Machine Learning Research*. 2022;23(47):1-33. http://jmlr.org/papers/v23/21-0635.html

30. Song J, Meng C, Ermon S. Denoising diffusion implicit models. *arXiv preprint arXiv:201002502*. Published online 2020.

31. Landman BA, Huang AJ, Gifford A, et al. Multi-parametric neuroimaging reproducibility: A 3-T resource study. *Neuroimage*. 2011;54(4):2854-2866. doi:https://doi.org/10.1016/j.neuroimage.2010.11.047

32. LaMontagne PJ, Benzinger TLS, Morris JC, et al. OASIS-3: Longitudinal Neuroimaging, Clinical, and Cognitive Dataset for Normal Aging and Alzheimer Disease. *medRxiv*. Published online January 1, 2019:2019.12.13.19014902. doi:10.1101/2019.12.13.19014902

33. Shock NW. *Normal Human Aging: The Baltimore Longitudinal Study of Aging*. US Department of Health and Human Services, Public Health Service, National …; 1984.

34. Bischoff-Grethe A, Ozyurt IB, Busa E, et al. A technique for the deidentification of structural brain MR images. *Hum Brain Mapp*. 2007;28(9):892-903. doi:https://doi.org/10.1002/hbm.20312

35. Gulban OF, Nielson D, lee john, et al. poldracklab/pydeface: PyDeface v2.0.2. *Zenodo*. Preprint posted online July 2022. doi:10.5281/zenodo.6856482

36. Schimke N, Hale J. Quickshear Defacing for Neuroimages. *HealthSec*. 2011;11:11.

37. Otsu N. A threshold selection method from gray-level histograms. *Automatica*. 1975;11(285-296):23-27.

38. Virtanen P, Gommers R, Oliphant TE, et al. SciPy 1.0: fundamental algorithms for scientific computing in Python. *Nat Methods*. 2020;17(3):261-272. doi:10.1038/s41592-019-0686-2

39. Maneewongvatana S, Mount DM. Analysis of approximate nearest neighbor searching with clustered point sets. *arXiv preprint cs/9901013*. Published online 1999.

40. Liu Y, Huo Y, Dewey B, Wei Y, Lyu I, Landman BA. Generalizing deep learning brain segmentation for skull removal and intracranial measurements. *Magn Reson Imaging*. 2022;88:44-52. doi:https://doi.org/10.1016/j.mri.2022.01.004

41. Kreilkamp BAK, Martin P, Bender B, et al. Big Field of View MRI T1w and FLAIR Template - NMRI225. *Sci Data*. 2023;10(1):211. doi:10.1038/s41597-023-02087-1

42. Yang Q, Yu X, Lee HH, et al. Label efficient segmentation of single slice thigh CT with two-stage pseudo labels. *Journal of Medical Imaging*. 2022;9(5):052405. doi:10.1117/1.JMI.9.5.052405

43. Yu X, Tang Y, Yang Q, et al. Accelerating 2D abdominal organ segmentation with active learning. In: *Proc.SPIE*. Vol 12032. 2022:120323F. doi:10.1117/12.2611595

44. Yu X, Tang Y, Yang Q, et al. Longitudinal variability analysis on low-dose abdominal CT with deep learning-based segmentation. In: *Proc.SPIE*. Vol 12464. 2023:1246423. doi:10.1117/12.2653762

45. Lindstrom MJ, Bates DM. Newton—Raphson and EM Algorithms for Linear Mixed-Effects Models for Repeated-Measures Data. *J Am Stat Assoc*. 1988;83(404):1014-1022. doi:10.1080/01621459.1988.10478693

46. He K, Zhang X, Ren S, Sun J. *Deep Residual Learning for Image Recognition*.; 2016. http://image-net.org/challenges/LSVRC/2015/

47. Chen S, Ma K, Zheng Y. Med3d: Transfer learning for 3d medical image analysis. *arXiv preprint arXiv:190400625*. Published online 2019.

48. Wang T, Zhang Y, Qi S, Zhao R, Xia Z, Weng J. Security and Privacy on Generative Data in AIGC: A Survey. *ACM Comput Surv*. 2024;57(4). doi:10.1145/3703626

49. Deng J, Guo J, Xue N, Zafeiriou S. ArcFace: Additive Angular Margin Loss for Deep Face Recognition. In: *Proceedings of the IEEE/CVF Conference on Computer Vision and Pattern Recognition (CVPR)*. 2019.

50. Ding Y, Wu G, Chen D, et al. DeepEDN: A Deep-Learning-Based Image Encryption and Decryption Network for Internet of Medical Things. *IEEE Internet Things J*. 2021;8(3):1504-1518. doi:10.1109/JIOT.2020.3012452

51. Thompson PM, Stein JL, Medland SE, et al. The ENIGMA Consortium: large-scale collaborative analyses of neuroimaging and genetic data. *Brain Imaging Behav*. 2014;8(2):153-182. doi:10.1007/s11682-013-9269-5

52. Baker BT, Abrol A, Silva RF, et al. Decentralized temporal independent component analysis: Leveraging fMRI data in collaborative settings. *Neuroimage*. 2019;186:557-569. doi:https://doi.org/10.1016/j.neuroimage.2018.10.072

53. Rieke N, Hancox J, Li W, et al. The future of digital health with federated learning. *NPJ Digit Med*. 2020;3(1):119. doi:10.1038/s41746-020-00323-1

54. Eke D, Aasebø IEJ, Akintoye S, et al. Pseudonymisation of neuroimages and data protection: Increasing access to data while retaining scientific utility. *Neuroimage: Reports*. 2021;1(4):100053. doi:https://doi.org/10.1016/j.ynirp.2021.100053

# Supplementary Materials

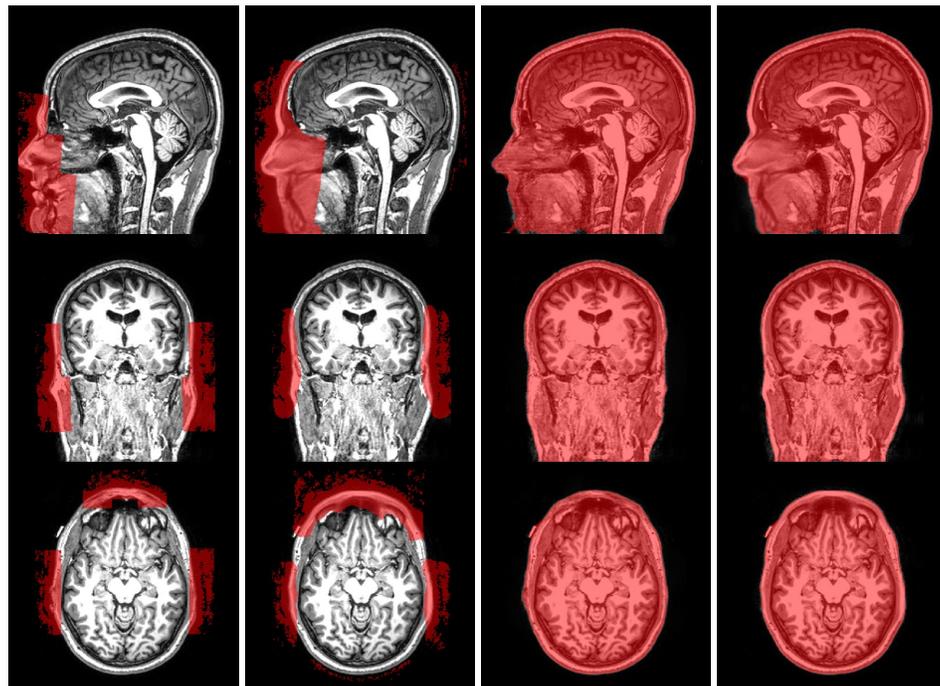

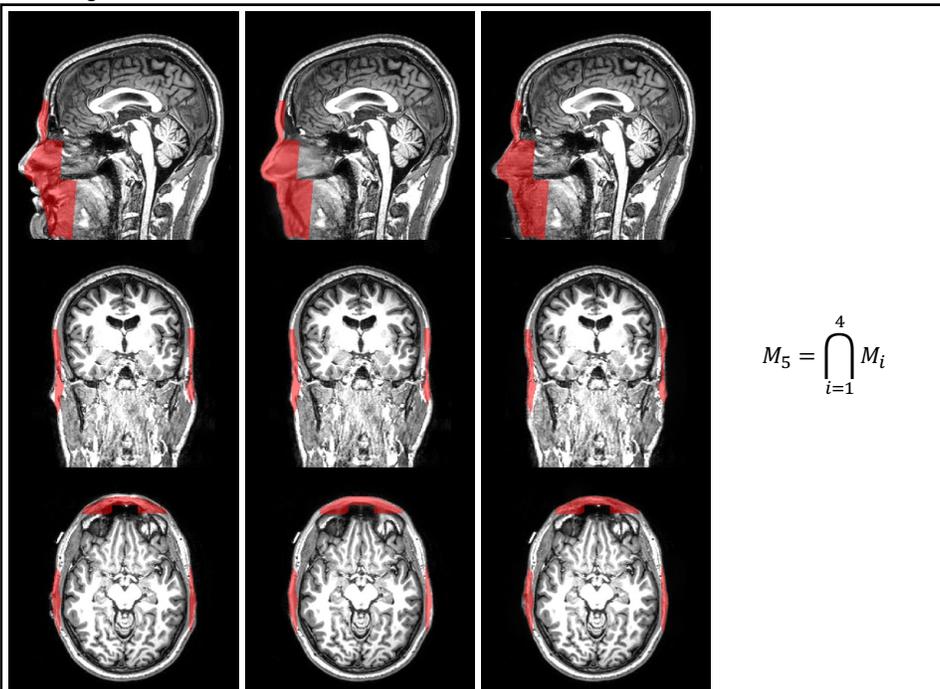

$$M_5 = \bigcap_{i=1}^{4} M_i$$

**Figure S1.** The face area, for which PSNR and SSIM were computed, is defined as the intersection of four masks.

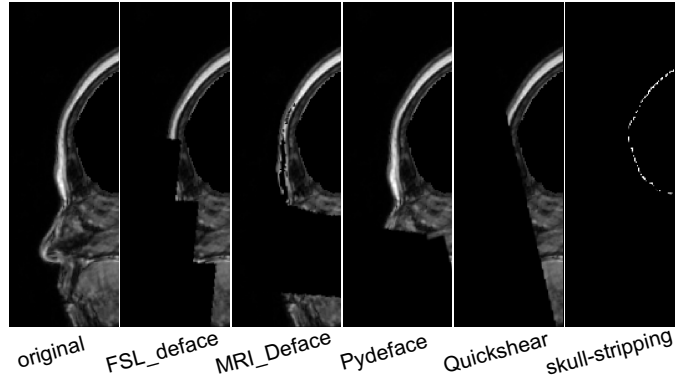

original  FSL_deface  MRI_Deface  Pydeface  Quickshear  skull-stripping

**Figure S2.** Six types of images are used for predicting abdomen, thigh, and shin muscle radiodensities from facial voxels in head MRIs. To focus on the facial region, the posterior two-thirds of each 3D volume have been cropped out, leaving only the anterior portion. Additionally, the brain has been masked out to further isolate the facial voxels. The resulting images have dimensions of 201*87*261 with an isotropic resolution of 1 mm$^3$. Here, we display the middle sagittal slice of each image type.